\documentclass[manuscript,screen]{acmart}
\usepackage{pdflscape}
\usepackage{subcaption}
\usepackage{longtable}
\usepackage{paralist}
\usepackage[inline]{enumitem}
\AtBeginDocument{%
  }

\setcopyright{acmlicensed}
\copyrightyear{2018}
\acmYear{2018}
\acmDOI{XXXXXXX.XXXXXXX}
\acmConference[Conference acronym 'XX]{Make sure to enter the correct
  conference title from your rights confirmation email}{June 03--05,
  2018}{Woodstock, NY}
\acmISBN{978-1-4503-XXXX-X/18/06}

\begin{document}

\title[A Closer Look at the Existing Risks of Generative AI]{A Closer Look at the Existing Risks of Generative AI: \\ Mapping the Who, What, and How of Real-World Incidents}

\author{Megan Li}
\affiliation{%
  \institution{Carnegie Mellon University}
  \city{Pittsburgh}
  \country{USA}}
\email{meganli@andrew.cmu.edu}

\author{Wendy Bickersteth}
\affiliation{%
  \institution{Carnegie Mellon University}
  \city{Pittsburgh}
  \country{USA}}
\email{wbickers@andrew.cmu.edu}

\author{Ningjing Tang}
\affiliation{%
  \institution{Carnegie Mellon University}
  \city{Pittsburgh}
  \country{USA}}
\email{ningjingt@andrew.cmu.edu}

\author{Jason Hong}
\affiliation{%
  \institution{Carnegie Mellon University}
  \city{Pittsburgh}
  \country{USA}}
\email{jasonh@cs.cmu.edu}

\author{Lorrie Cranor*}
\affiliation{%
  \institution{Carnegie Mellon University}
  \city{Pittsburgh}
  \country{USA}}
\email{lorrie@cmu.edu}

\author{Hong Shen*}
\affiliation{%
  \institution{Carnegie Mellon University}
  \city{Pittsburgh}
  \country{USA}}
\email{hongs@andrew.cmu.edu}

\author{Hoda Heidari*}
\affiliation{%
  \institution{Carnegie Mellon University}
  \city{Pittsburgh}
  \country{USA}}
\email{hheidari@cmu.edu}

\authorsaddresses{}

\renewcommand{\shortauthors}{Li et al.}

\begin{abstract}
Due to its general-purpose nature, Generative AI is applied in an ever-growing set of domains and tasks, leading to an expanding set of risks of harm impacting people, communities, society, and the environment. These risks may arise due to failures during the design and development of the technology, as well as during its release, deployment, or downstream usages and appropriations of its outputs. In this paper, building on prior taxonomies of AI risks, harms, and failures, we construct a taxonomy specifically for \emph{Generative} AI failures and map them to the harms they precipitate. Through a systematic analysis of 499 publicly reported incidents, we describe \textit{what} harms are reported, \textit{how} they arose, and \textit{who} they impact. We report the prevalence of each type of harm, underlying failure mode, and harmed stakeholder, as well as their common co-occurrences. We find that most reported incidents are caused by \textit{use-related issues} but bring harm to parties \textit{beyond} the end user(s) of the Generative AI system at fault, and that the landscape of Generative AI harms is distinct from that of traditional AI. Our work
offers actionable insights to policymakers, developers, and Generative AI users.
In particular, we call for the prioritization of non-technical risk and harm mitigation strategies, including public disclosures and education and careful regulatory stances. 
\end{abstract}


\received{27 May 2025}

\maketitle
\footnotetext[1]{$^{*}$Co-last authors contributed equally to this work.}

%

\section{Introduction} \label{introduction}

Generative AI -- defined as AI that produces novel output in the form of text, images, audio, or video \cite{weidinger_sociotechnical_2023} -- has taken the world by storm, presenting both risks and benefits to humanity. While there is broad agreement on the importance of mitigating the risks of Generative AI, there is a widening divide between those who emphasize its existential risks 
and those more focused on its existing risks. \citet{bostrom} defines existential risks as ``one where an adverse outcome would either annihilate Earth-originating intelligent life or permanently and drastically curtail its potential.'' We define existing risk as ``one whose instances have been observed in the real world and caused harm to individuals, communities, organizations, or the environment.'' 

Although some investment in assessing the existential risks of AI is essential for long-term planning \cite{hendrycks_overview_2023,karger2023forecasting, kasirzadeh_two_2025}, effective risk mitigation today requires understanding how harms from Generative AI technologies currently emerge. In this paper, we focus on characterizing the existing risks of Generative AI. We argue that addressing these risks is critical not only because they have already caused harm and are likely to persist, but also because they can accumulate over time into more severe harms (e.g., to the environment and democracy).

The AI Risk Management Framework (AI RMF) developed by the U.S. National Institute of Standards and Technology (NIST) emphasizes risk \emph{identification} and \emph{prioritization} as critical steps in addressing AI risks \cite{ai2023artificial}. A growing body of literature identifies the risks and harms of AI across various domains and applications (see Table \ref{tab:tax_overview} in the Appendix for an overview of prior work). However, much of this work lacks grounding in real-world evidence and often fails to address \emph{who} is affected and \emph{how} the risk or harm arises, limiting its effectiveness in guiding risk prioritization and mitigation. In this work, we aim to fill this gap by providing a large-scale, empirical mapping of real-world incidents of harm caused by Generative AI systems, tracing each harm from the underlying sociotechnical issue(s) or \emph{failure(s)} that precipitated it to the downstream impact it had on people and society. 

The AI RMF recommends prioritizing risks based on their \textit{projected impact} \cite{ai2023artificial}. While assessing projected impact is inherently complex, two key factors should guide this process: 
\begin{enumerate*}
    \item the \emph{prevalence} of the risk -- how frequently it causes harm in practice -- and
    \item \emph{who} is at risk. 
\end{enumerate*} 
This motivates our first two research questions. \textbf{RQ1:} What is the nature and prevalence of Generative AI harms observed in practice today? \textbf{RQ2:} Who is most at risk of these harms? 
To effectively inform risk mitigation strategies, it is imperative to identify the underlying issue(s) -- which we refer to as \textit{sociotechnical failure modes} (Section \ref{sec:failure_modes}) -- that led to the incident of harm. While prior work has proposed frameworks to identify where technology fails \cite{raji_fallacy_2022, macrae2022learning, slattery_ai_nodate, fmeas, stpa}, much of this work is focused on technical failures, is not grounded in real-world evidence, and does not consider how \emph{Generative} AI may induce a unique landscape of failures and risks compared to other AI technologies. Identifying Generative AI failure modes and mapping them to real-world harms can guide more effective mitigation strategies, motivating our third research question. \textbf{RQ3:} What is the nature and prevalence of \textit{sociotechnical failure modes} leading to existing harms of Generative AI?

To ground our investigation of these questions in real-world evidence, we follow recent work that uses repositories of publicly reported AI incidents as a critical source of empirical data for mapping and measuring AI harms \cite{lee_deepfakes_2024, abercrombie_collaborative_2024, pittaras_taxonomic_nodate, hoffman, raji_fallacy_2022, hutiri_not_2024}. We performed a systematic analysis of 499 publicly reported Generative AI incidents, identifying \textit{what} harms occurred, \textit{how} they materialized, and \textit{who} experienced them. Through our analysis, we validated and revised prior taxonomies of AI harms (Figure \ref{fig:risk_tax}) and developed a novel taxonomy of sociotechnical failure modes (Figure \ref{fig:failure_tax}), which we present alongside a quantitative analysis of how they relate to each other. Importantly, we find that most incidents in our analysis bring harm to stakeholders that did \textit{not} interact directly with the Generative AI system, undermining the accountability frameworks assumed in many AI governance models. We also find malicious use to be a prevalent failure mode underlying Generative AI harms, in contrast to previous work that analyzes AI incidents more broadly~\cite{velazquez_decoding_2024}.

It is important to note that our dataset is likely not representative of real-world Generative AI harms in terms of scope or frequency. In particular, databases of publicly reported incidents may disguise the actual prevalence of harms due to factors such as reporting bias. 
Nonetheless, we believe our study provides essential empirical evidence to ground discussions of AI risk management, helping researchers and policymakers prioritize existing risks, understand where harm mitigation efforts are most urgently needed, and how \textit{Generative} AI requires novel risk mitigation strategies (Section \ref{discussion}).


\section{Related Work} \label{related_work}

In this paper, we define 
harm as ``an outcome negatively affecting people and their interests" and risk as ``the magnitude and likelihood of harm" \cite{ai2023artificial}. We view Generative AI risks and harms in the sociotechnical contexts where they arise; that is, we consider not only  technical but also human and societal factors to play critical roles in precipitating positive or negative outcomes associated with Generative AI \cite{weidinger_sociotechnical_2023, shelby_sociotechnical_2023}. 

\subsection{Taxonomies of AI Risks and Harms}

Risk and harm taxonomies provide a common vocabulary through which stakeholders can report, evaluate, and mitigate system failures \cite{koessler_risk_2023}. 
There is some consensus on the broad domains of risk associated with AI. These include risks to the physical or psychological well-being of people, human rights and civil liberties, political and economic structures, society and culture, and the environment \cite{abercrombie_collaborative_2024}. 
Some taxonomies also consider multiple dimensions of harm, such as the level of severity, the scope of the impacted entities, and  whether the harm is reversible \cite{oecd}. For example, \citet{hoffman} proposed a two-dimensional taxonomy of AI harms with domain on one axis and level of realization on the other. Using this taxonomy, harm is described by both a domain category such as \textit{Physical Health/Safety} or \textit{Privacy} and whether the harm was realized or yet unrealized. Table \ref{tab:tax_overview} (in the Appendix) provides a nonexhaustive overview of these taxonomies, summarizing the technologies they apply to, their target audiences, and the methodology of their work. 

Throughout this paper, we build on Weidinger et al.'s highly-cited risk taxonomy for Generative AI systems, developed through a literature review, and Abercrombie et al.'s harm taxonomy for AI and automated systems, developed through a combination of use case analysis, literature review, and annotation testing on publicly reported incidents \cite{weidinger_sociotechnical_2023, abercrombie_collaborative_2024}. Our contribution goes beyond existing risk and harm taxonomies in two ways: first, we report the quantitative prevalence of different Generative AI harms based on real-world incidents, and second, we map harms to the stakeholders they affect and the underlying issues that precipitated them.

\subsection{Sociotechnical Failure Modes}
\label{sec:failure_modes}
The harms of Generative AI do not materialize in a vacuum; instead, they arise due to a variety of human and technical factors and their interactions \cite{macrae2022learning}. To describe these underlying sources of harm, we introduce the term \textit{sociotechnical failure modes}, which we define as ``the ways in which a technical system, its creators or users, or the various societal structures interacting with it bring risk or harm to some stakeholders.'' Most AI risk and harm taxonomies do not address the topic of \textit{how} risks and harms surface \cite{turrietal}.
As steps toward filling this gap, \citet{raji_fallacy_2022} introduce a taxonomy of AI system failures and \citet{slattery_ai_nodate} organize AI risks in a \textit{causal} taxonomy.

The term ``failure mode" is used in safety engineering contexts such as Failure Mode and Effects Analysis (FMEA), a framework for identifying and prioritizing failure modes by \begin{enumerate*}
    \item listing out ``the functions of a component/system or steps of a process," 
    \item identifying the failure modes --- the ``mechanisms by which each function or step can go wrong," and
    \item identifying the \textit{impact} and \textit{cause} of each failure mode \cite{fmeas}.
\end{enumerate*}
Recent work proposes the application of safety engineering approaches including FMEA 
to assess and mitigate the social and ethical risks of machine learning systems \cite{rismani_SafetyEng}. However, our method of identifying failure modes using incident reports
is distinct from FMEA in two ways: first, it is performed after the failure has occurred, making it more akin to a postmortem risk assessment; and second, because we are limited to publicly available data, we cannot examine the specific components, functions, or control actions of the system. 

Another safety engineering framework that may be applicable to machine learning systems is System Theoretic Process Analysis (STPA), which involves mapping the components of a system and their interactions to identify potential sources of harm \cite{stpa}. As \citet{rismani_SafetyEng} note, STPA is better suited for capturing emergent phenomena rather than individual component behavior, congruent with increasing work toward thinking about AI failures in sociotechnical contexts \cite{tangfailurecards, weidinger_sociotechnical_2023, shelby_sociotechnical_2023}. While our analysis in this work is less structured than what would constitute an application of STPA, we draw inspiration from the first two steps of the framework. First, we ``[identify] losses via outlining stakeholders and their values" by identifying harms via first describing \textit{who} is affected in each incident. Second, we ``model the \textit{control structure} of the full sociotechnical system" by situating failures within the three-layered framework introduced in \citet{weidinger_sociotechnical_2023} 
to describe the sociotechnical system of Generative AI composed of \textit{Capability}, \textit{Human interaction}, and \textit{Systemic impact} layers. 



\subsection{AI Incidents as Empirical Data}

Prior work on identifying AI harms and failures highlights repositories of publicly reported AI incidents as sources of empirical data \cite{lee_deepfakes_2024, abercrombie_collaborative_2024, pittaras_taxonomic_nodate, hoffman, raji_fallacy_2022, hutiri_not_2024, velazquez_decoding_2024}. For example, \citet{lee_deepfakes_2024} systematically reviewed 321 publicly reported incidents to develop a taxonomy of AI privacy risks. Similarly, \citet{raji_fallacy_2022} reviewed 283 publicly reported incidents to develop their taxonomy of AI system failures.  

Three commonly cited repositories are the AI Incident Database (AIID); the OECD AI Incidents Monitor; and the AI, Algorithmic and Automation Incidents and Controversies (AIAAIC) repository. All of these repositories maintain harm taxonomies for incident annotation and categorization \cite{aiid, aim, aiaaic_repo}. Each repository also provides their own definition of an \textit{incident}; for example, the AIAAIC's definition is ``a sudden known or unknown event (or `trigger') that becomes public and which takes the form of a disruption, loss, emergency, or crisis'' \cite{aiaaic_repo}. 

Although the reported-incident analysis approach has several key limitations due to factors including reporting bias (further discussed in Section \ref{limitations}), the assumption underpinning this approach is that these incidents reflect real-world harms arising in sociotechnical systems and hence serve as reasonable raw material on which to test and apply risk mapping frameworks. Incident repositories also play an important role in reducing risk in other safety-critical domains such as aviation and cybersecurity \cite{turrietal}. 

\subsection{How, What, Who: Mapping Failures to Harms to Stakeholders}
\label{how-what-who}
\citet{slattery_ai_nodate} introduced a causal taxonomy of AI risks where each risk is mapped to the \textit{entity} (human, AI, or other) that precipitated it, that entity's \textit{intent}, and \textit{when} (pre-deployment, post-deployment, or other) the risk arose. Their work contributes to an understanding of the sociotechnical factors that precipitate AI risks and harms, which may help develop novel harm mitigation strategies and identify the stakeholders best positioned to address specific risks. However, in addition to identifying only coarse-grained causal factors, their work does not identify \textit{who} tends to experience each type of risk or harm, limiting its utility to prioritize risks and map accountability pathways. 



To address this gap, \citet{velazquez_decoding_2024} applied a ``what-where-who" framework to analyze 639 AI incident reports. They used large language models to operationalize their framework and automatically classify each incident based on \textit{what} type of harm occurred, \textit{where} it originated, and \textit{who} was harmed. They found that the vast majority of incidents in their dataset harmed the individual directly interacting with the AI system. \citet{hutiri_not_2024} arrived at a similar conceptual framework to characterize the harms of speech generators, mapping harms from their \textit{responsible} to \textit{affected entities}. However, they found that most harms did \textit{not} affect the individual who interacted with the speech generator. Taken together, these papers highlight the importance of mapping harms to the stakeholders they affect. They also suggest that there are critical differences between traditional AI harms and Generative AI harms in terms of who they affect and how they materialize.

In this work, we build on these insights to characterize the how, what, and who of Generative AI harms through a systematic manual analysis of 499 publicly reported incidents. We report the quantitative prevalence of the harms we identify (the \emph{what}), the failure modes that precipitate them (the \emph{how}), and the stakeholders who were harmed (the \emph{who}). By grounding our analysis in real-world incidents, our work offers a novel, empirical understanding of the sociotechnical factors that contribute to  Generative AI harms. Our results show how the landscape of Generative AI harms is distinct from traditional AI harms, reveal where pathways of accountability break down, and emphasize the potential impact of non-technical harm mitigation.   



\section{Methods} \label{methods}

\subsection{Using Case Studies to Construct Generative AI Risk and Failure Mode Taxonomies}
To develop taxonomies of Generative AI harms and sociotechnical failure modes exhibited in publicly reported incidents, we performed a systematic review of Generative AI incident reports. We framed each incident as a case study, where the goal was to determine \begin{enumerate*}
    \item the output \textit{modality} of the Generative AI system at fault,
    \item the \textit{human stakeholders} who experienced harm,
    \item the \textit{nature} of the harm experienced, and
    \item any and all \textit{sociotechnical failure modes} that contributed to the harm.
\end{enumerate*} 
Table \ref{tab:annotation_cols} provides further details for each of these four attributes. The third and fourth attributes are the building blocks for our taxonomies of 
harms and 
sociotechnical failure modes, respectively.

\begin{table*}[h!]
    \centering
    \caption{Summary of Coding Columns}
    {\begin{tabular}{p{1.25in} p{2.5in} p{1.75in}}\hline
        \textbf{Column} & \textbf{Description} & \textbf{Example entries} \\\midrule
        Modality & The output modality of the Generative AI system. & text, image, audio, video \\ \hline
        Affected stakeholders (up to 2) & The human stakeholders that suffered actual or worst-case scenario harm as a result of the Generative AI system.	& end user, individual(s) beyond the end user, community, society \\ \hline
        Harm (up to 2) & The actual or worse-case scenario harm experienced by the affected stakeholders. & Impersonation/identity theft, Pollution of information ecosystem \\ \hline
        Sociotechnical failure mode (all that apply) & The technical or user issue that most proximally contributed to the harm. & Hallucination, Malicious use, Failure of safety guardrails
    \end{tabular}}
    \label{tab:annotation_cols}
\end{table*}

We sourced incident reports from both the AIID \cite{aiid} and the AIAAIC \cite{aiaaic_repo}. Both sources include incidents that involve a wide range of AI systems, so we manually filtered the union of the two databases to include only incidents where a Generative AI was at fault, resulting in our dataset of 499 Generative AI incidents. We did not use the OECD AI Incidents Monitor \cite{aim} because it is automatically curated (whereas the incidents in the AIID and AIAAIC undergo manual review) and its definition of ``incident" is broader than those used in the other two repositories. We scoped our study around incidents -- where ``incident” is defined as a single event -- that have been manually submitted and reviewed. 

We based our analysis of both harms and sociotechnical failure modes on previous literature and used an iterative coding process to categorize each of the four attributes described in Table \ref{tab:annotation_cols}. As a preliminary codebook for our analysis of Generative AI harms, we constructed a union of previous harm taxonomies developed by  \citet{abercrombie_collaborative_2024} and \citet{weidinger_sociotechnical_2023}. As a preliminary codebook for Generative AI sociotechnical failure modes, we used Raji et al.'s taxonomy of AI system failures as a baseline and used Slattery et al.'s causal taxonomy of AI risks as a framework within which to brainstorm additional possible failure modes \cite{raji_fallacy_2022, slattery_ai_nodate}. We then expanded and refined each of these codebooks through a systematic review of our incident dataset.  
Our codebooks are available in the Appendix (see Sections \ref{appendix_harm}-\ref{appendix_entities}). 

\paragraph{Significant revisions to the codebook of harms} Both \citet{weidinger_sociotechnical_2023} and \citet{abercrombie_collaborative_2024} consider privacy violations to be a specific harm rather than a broad category. Therefore, we introduced a new category of \textit{Privacy \& Security} harms and populated it using the taxonomy of AI privacy risks introduced in \citet{lee_deepfakes_2024}. We also surfaced one new harm, \textit{Overburdening ecosystems}, not captured in our review of prior literature. In total, our taxonomy of Generative AI harms has 41 harms organized into 12 categories (Figure \ref{fig:risk_tax}).

\paragraph{Significant revisions to the codebook of sociotechnical failure modes} We first revised each AI system failure described in \citet{raji_fallacy_2022} to be more specific to Generative AI. For example, we increased the granularity of Implementation Failures by deriving three failure modes from its descriptions of implementation issues: \textit{Hallucination}, \textit{Failure of commonsense reasoning}, and \textit{Failure to comply with contextual norms}. We also noted that Raji et al.'s work is specifically concerned with engineering failures, but there were a large number of incidents with use-related failure modes in our dataset. Thus, we generated additional failure modes based on themes that arose in the data and informed by prior literature \cite{walkowiak_generative_2024}. In total, our taxonomy of Generative AI failure modes has 14 failure modes organized into 4 categories (Figure \ref{fig:failure_tax}).

\subsection{Qualitative Analysis Procedure}
The first and second authors collaborated to code the incident database with the four attributes summarized in Table \ref{tab:annotation_cols}. After developing the preliminary codebooks, the two authors convened for a training phase on a random 10\% of the data. They then independently coded the remainder of the dataset, reconvening periodically to compare codes, discuss, and reach consensus for all disagreements.

The analysis was framed by three questions inspired by \citet{slattery_ai_nodate} and \citet{hutiri_not_2024}:
\begin{enumerate*}
    \item Which human entities are mentioned in the incident summary?
    \item Which entities were harmed and what is the nature of the harm? 
    \item Was the most proximal source of risk or harm primarily human or technical in nature? Was the harm intentional or unintentional? Did the risk materialize pre- or post-deployment?
\end{enumerate*}

In cases where there was insufficient information to determine whether actual harm occurred, the authors imagined the ``worst-case scenario.'' For example, in one incident, a chatbot convinced its user to euthanize their dog \cite{dog_AIAAIC1751}. In the worst-case scenario, this choice was incorrect and the user suffered significant harm. The authors discussed this incident as if this worst-case scenario were true and coded it as such. Additionally, the coders chose the two most \textit{severe} harms in cases where there were more than two types of harm, where severity was measured by the scale of the harmed stakeholder and the magnitude of possible harm. 
The complete coded incident dataset is available in Section \ref{appendix_incidents} in the Appendix.

\subsection{Limitations} \label{limitations}

As with other work that relies primarily on databases of publicly reported incidents for empirical data, this paper has methodological limitations.

\paragraph{Some harms can have a cumulative nature} Both databases from which we sourced incident reports define an \textit{incident} as a single \textit{event} (the AIAAIC further specifies it to be ``sudden") \cite{aiid, aiaaic_repo}. This means that harms that do not have single precipitating events, but instead accumulate over time, are unlikely to surface in these databases. For example, many harms to society (e.g., the devaluation and loss of human creativity and critical thinking) or the environment (e.g., global warming) can only be measured over the course of months or years. While it is possible that certain incidents are indicative of cumulative harms -- for example, Generative AI systems trained on copyrighted art may produce copyright violations today but lead to the devaluation of human artists over time -- we intentionally avoided this kind of speculation in our coding process as we acknowledge that the landscape of Generative AI changes rapidly. Instead, we tried to be comprehensive in our consideration of shorter-term harms. 

\paragraph{Harms may be over- or under-represented due to reporting bias} Any database of publicly reported incidents will reflect some level of reporting bias. Concepts that are sensational or relatively well-understood by the public, such as deepfake impersonations, may be overreported. Concepts related to the design or development of Generative AI, such as data acquisition and cleaning or the implementation of safety guardrails, may be underreported, perhaps in part because industry is not incentivized or required to publicly report issues they surface during the design, development, and testing of their products. 
Additionally, the unsanctioned use of Generative AI in academic or workplace settings may be intentionally concealed and therefore infrequently reported despite high incidence in the real world. 

Still, incident repositories serve as important starting points from which to map and estimate the frequency of materialized harms. Future partnerships with institutions such as MITRE, which is developing a confidential incident repository, may improve representativeness. 


\section{Findings}  \label{findings}

\begin{figure}[h]
    \centering   
    \includegraphics[width=1\linewidth]{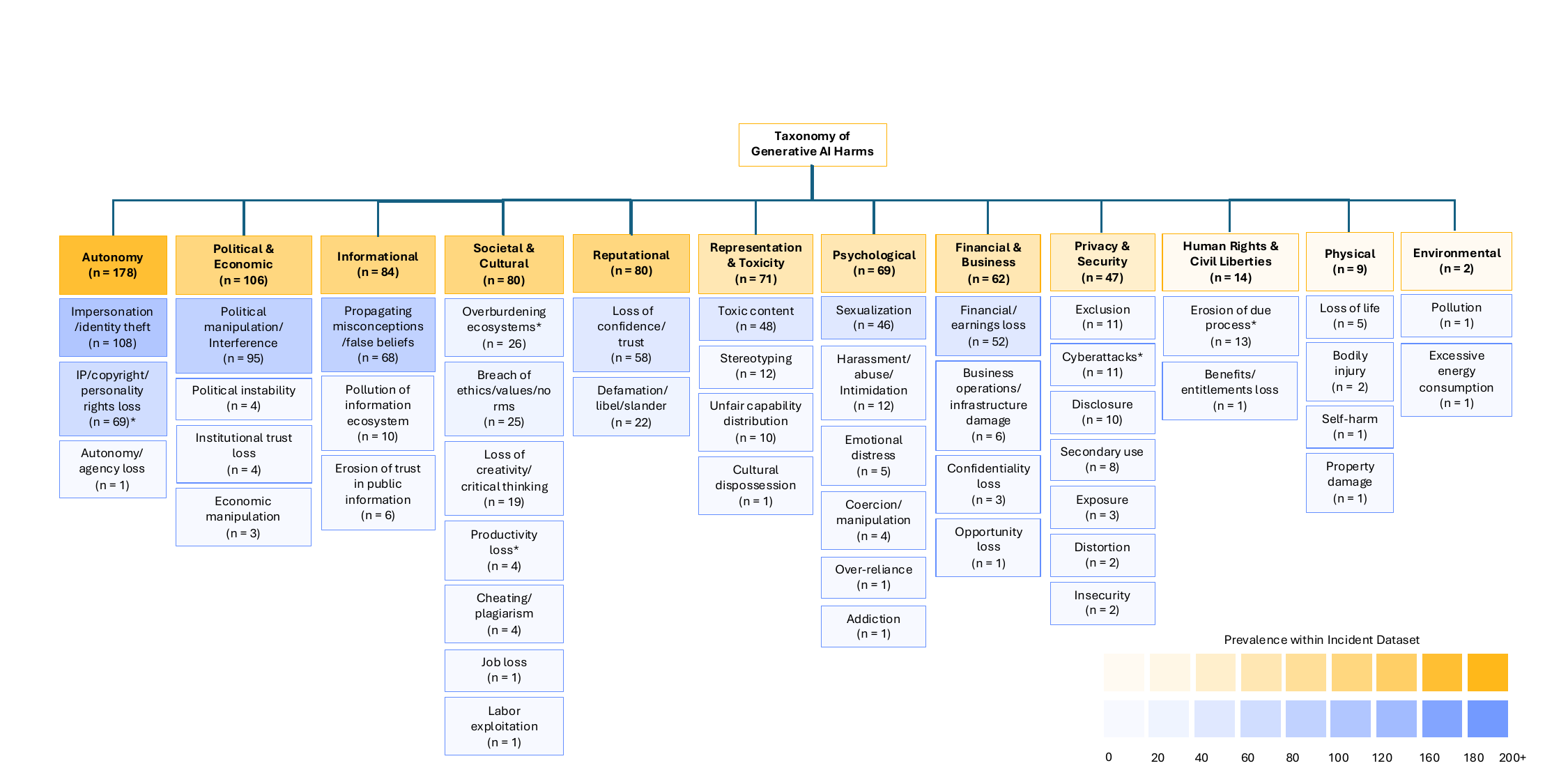}
    \caption{Overview of Taxonomy of Generative AI Harms. Darker colors indicate higher prevalence in our dataset. Harms that were heavily altered or created are indicated by an asterisk.}
    \label{fig:risk_tax}
\end{figure}

We report the Generative AI harms (Figure \ref{fig:risk_tax}) and sociotechnical failure modes (Figure \ref{fig:failure_tax}) surfaced in our incident dataset, their prevalence, who they affect, and how they relate to each other.\footnote{Harms and failure modes that are included in our complete codebooks (see Sections \ref{appendix_harm} and \ref{appendix_failures}) but were not applied during the coding process are omitted here.} We also investigate whether particular output modalities are more often associated with certain types of harms and failure modes. While most of the harms we identified are well-defined and taxonomized in prior work \cite{weidinger_sociotechnical_2023, abercrombie_collaborative_2024, lee_deepfakes_2024}, we introduce our taxonomy of sociotechnical failure modes, as well as our investigation of the relationships between failure modes, harms, and who they affect, as novel contributions. When reporting prevalence data, we report the \textit{number of occurrences} of each harm or failure mode in our incident dataset. Because incidents were coded with up to two harms and as many failure modes as applicable, totals will be greater than the number of incidents.

\subsection{What Generative AI harms surface in publicly reported incidents?}

Our taxonomy of Generative AI harms surfaced in our incident dataset is made up of 41 harms, each falling into exactly one high-level category (Figure \ref{fig:risk_tax}). These harms and categories are largely identified and defined in previous literature \cite{abercrombie_collaborative_2024, weidinger_sociotechnical_2023, lee_deepfakes_2024}. 

We surfaced one new harm, \textit{Overburdening ecosystems}, and define it as ``the pollution of a space or ecosystem intended for human productivity or creativity (e.g., creative material submission or job application portals) with Generative AI.'' This harm is closely associated with risks to human creativity and critical thinking described in prior literature \cite{abercrombie_collaborative_2024}. However, previous literature focuses on the cumulative societal implications of automating human labor and creativity, while Overburdening ecosystems captures the short-term impacts of synthetic content -- often amounting to AI \textit{slop} -- on the individuals tasked with distinguishing AI from human. For example, several magazines have reported an influx of ``low-standard AI-generated'' submissions, overwhelming their editors and leading them to close their online submission portals \cite{submission_562}. 

\subsection{Who experiences the harms?}

To categorize \textit{who} experiences the harms of Generative AI, we define two types of stakeholders: \textit{interacting} 
stakeholders,
who directly and intentionally interact with the Generative AI system, and \textit{non-interacting} 
stakeholders, who do \textit{not} directly or intentionally interact with the Generative AI system. We further specify three interacting stakeholders: \begin{enumerate}
    \item \textit{end user (individual)}, referring to an end user of a Generative AI system who is an individual acting on their own interests;
    \item \textit{end user (organization)}, referring to an end user of a Generative AI system who is acting as or on behalf of an organization (e.g., a business);
    \item \textit{developer/deployer}, referring to the creators of a Generative AI system, including those developing foundation models and those carrying out downstream fine-tuning or other customizations;
\end{enumerate}
and four non-interacting stakeholders:
\begin{enumerate}
    \item \textit{individual(s) beyond the end user}, referring to individuals who are passive or third parties in the incident (e.g., the unknowing subject of a deepfake image);
    \item \textit{organization(s) beyond the end user}, referring to organizations that are passive or third parties in the incident (e.g., a magazine unknowingly receiving AI-generated submissions);
    \item \textit{community}, referring to a group of people with a common interest, culture, or within a small geographic region (e.g., visual artists); and
    \item \textit{society}, referring to a large or disparate aggregate of people and their institutions.
\end{enumerate}

\begin{figure}[h]
    \centering   \includegraphics[width=0.75\linewidth]{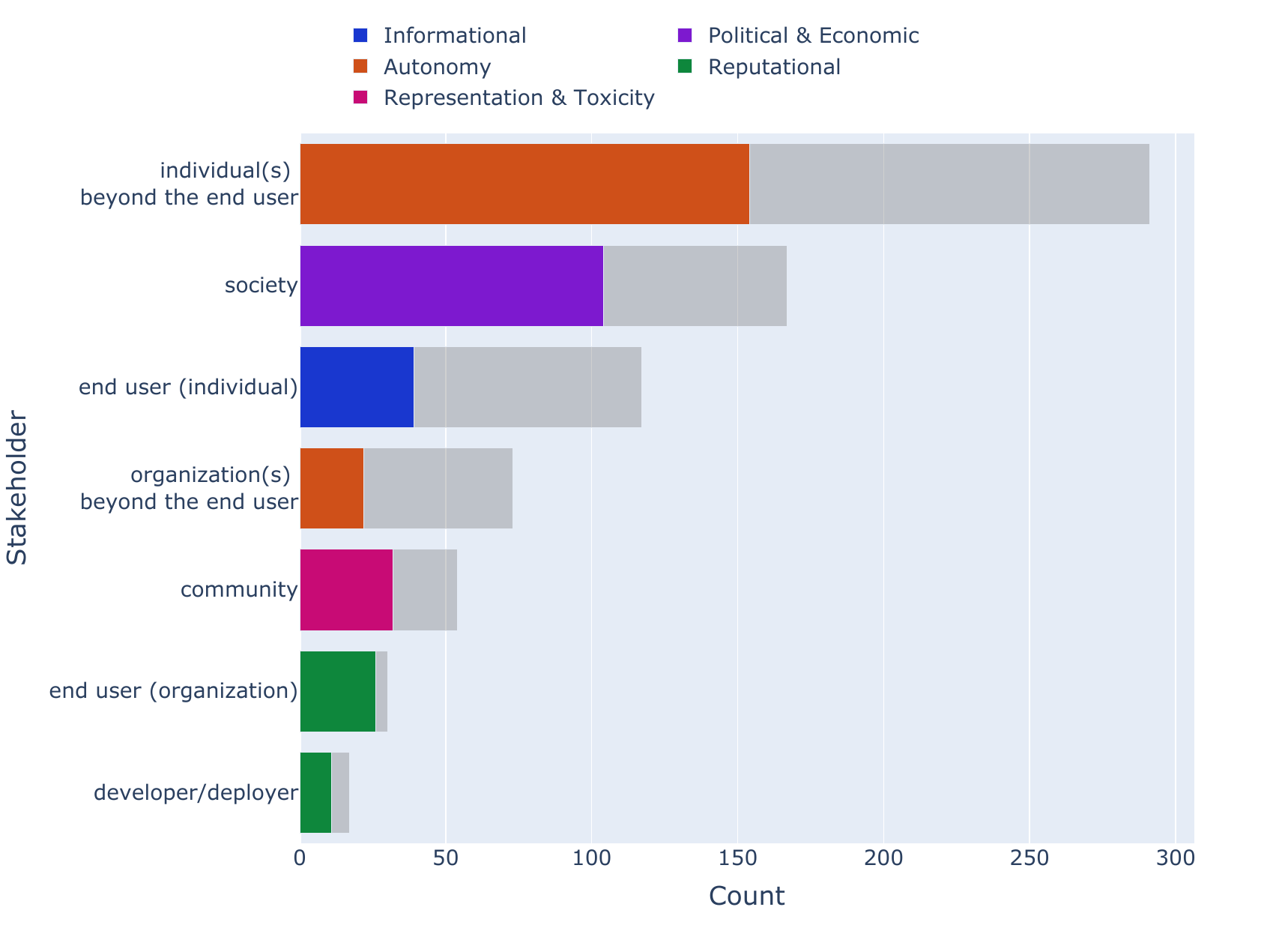}
    \caption{Distribution of harmed stakeholders, with the most prevalent category of harm for each stakeholder represented by the left (colored) side of the respective bar. The right (gray) side of each bar is all other categories of harm combined. The majority of incidents in our dataset harmed non-interacting stakeholders.}
    \label{fig:risk_cats}
\end{figure}

Figure \ref{fig:risk_cats} illustrates the distribution of who was affected and highlights the category of harm that most frequently affected each type of stakeholder. Most publicly reported incidents bring harm to non-interacting stakeholders. For example, Autonomy harms, the most prevalent category of harm in our dataset, accrued to individuals or organizations beyond the end user of the Generative AI system in all but one case. Similarly, Political \& Economic harms affected society in all but one case. The only category of risk that tended to affect interacting stakeholders was Reputational harms. 

These results suggest that stakeholders who tend to experience harm (third-party individuals and organizations, communities, and society) are not necessarily the same stakeholders thought to experience the benefits of Generative AI (the end users and creators). This represents a breakdown of one mechanism of accountability, whereby those \textit{choosing} to participate in the benefits of a technology must also assume the associated risks of harm. Our results suggest that those choosing to participate in the benefits of Generative AI can do so while subjecting themselves to relatively little risk, while those who did \textit{not} choose to participate experience the vast majority of harm while being deprived of the possible benefits. 




\subsection{How do Generative AI harms surface in publicly reported incidents?}

\begin{figure*}[h]
    \centering       \includegraphics[width=1\linewidth]{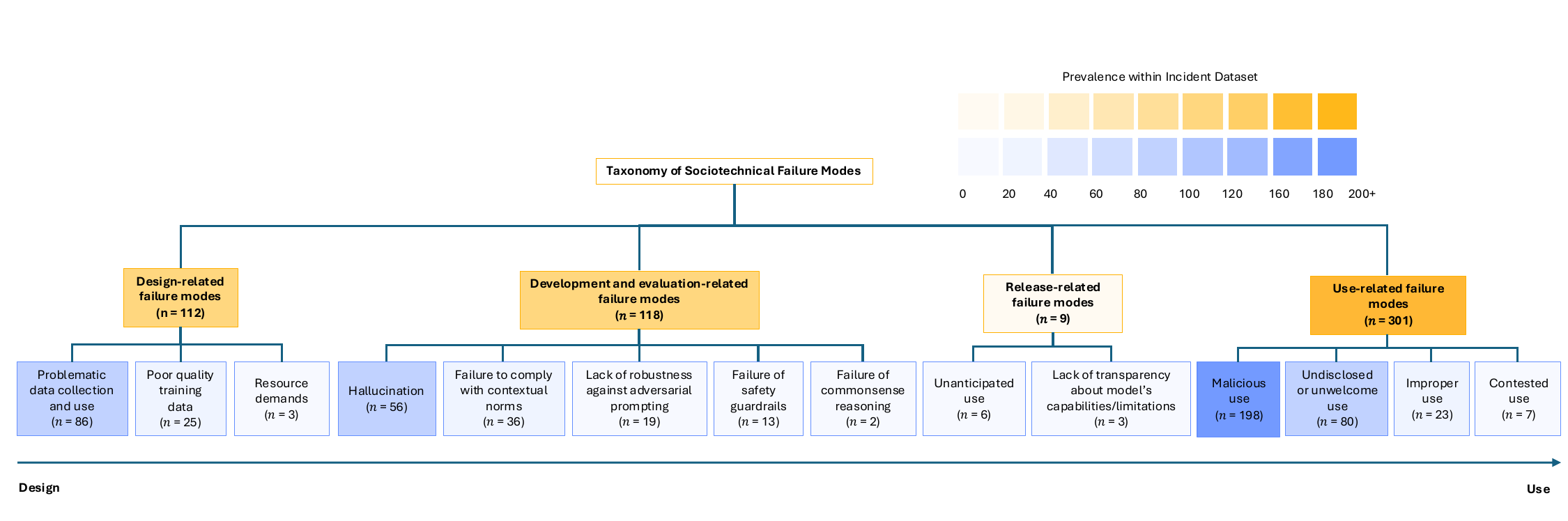}
    \caption{Overview of Taxonomy of Sociotechnical Failure Modes. Darker colors indicate higher overall prevalence in our dataset. 
    Use-related failure modes are the most prevalent of the four categories.}
    \label{fig:failure_tax}
\end{figure*}

As argued in Section \ref{sec:failure_modes}, the real-world harms of Generative AI are precipitated by a variety of sociotechnical failures. Although we acknowledge that we often cannot infer causal relationships between failures and harms based on the information available for incidents in our dataset, our analysis offers a high-level idea of what failure modes exist and are prevalent, and what kinds of harm they tend to precipitate.

We organize the 14 Generative AI sociotechnical failure modes we identified in our dataset based on whether they arise due to the \textit{design}, \textit{development and evaluation}, \textit{release}, or \textit{use} of Generative AI. Certain failure modes may in practice arise due to more than one of these stages, but we define each failure mode narrowly and classify it as accurately as possible. Furthermore, we situate our taxonomy of failure modes within Weidinger et al.'s sociotechnical framework for Generative AI safety evaluation, composed of Capability, Human interaction, and Systemic impact layers \cite{weidinger_sociotechnical_2023}.

\paragraph{Design-related failure modes ($n = 112$)}
These are failure modes arising due to the \textit{design} of a Generative AI system and cannot be addressed with safety features or improved model implementations. These exist largely at the Systemic impact layer, as they have to do with the broader systems into which the Generative AI system is integrated and are present regardless of the system's capabilities or human-AI interactions. We identify three design-related failure modes: \begin{enumerate*}
    \item \textit{Problematic data collection and use} ($n = 86$), referring to problems arising due to the indiscriminate nature of a system developers' collection and use of data, including preventing subjects (e.g., internet users) from giving their informed consent on how their data is used,
    \item \textit{Poor quality training data} ($n = 25$), referring to when the training data for a system reflects or exacerbates societal biases, overrepresents a certain type of content, or is polluted with poor quality information which is not sufficiently cleaned, and
    \item \label{Resource demands} \textit{Resource demands} ($n = 3$), referring to the high cost of training and using Generative AI for inference in terms of both physical and human resources.
\end{enumerate*} 

\paragraph{Development and evaluation-related failure modes ($n = 118$)} These are failure modes arising due to the \textit{development and evaluation} of a Generative AI system and exist largely at the Capability layer: that is, they determine whether the Generative AI system and its technical components are likely to exhibit problematic behaviors. We identify five development and evaluation-failure modes: \begin{enumerate*}
    \item \textit{Hallucination} ($n = 56$), referring to when a system makes false, misleading, or inaccurate claims as if they were facts,
    \item \textit{Failure of commonsense reasoning} ($n = 2$), referring to a system producing erroneous outputs resulting from a lack of intrinsic logic or commonsense reasoning, 
    \item \textit{Failure of safety guardrails} ($n = 13$), referring to when a safety feature either produces a new problem or simply fails,
    \item \textit{Failure to comply with contextual norms} ($n = 36$), referring to when a system fails to meet user expectations (e.g., producing gibberish, parroting), and
    \item \textit{Lack of robustness against adversarial prompting} ($n = 19$), referring to when a system misbehaves as a result of adversarial prompting.
\end{enumerate*}

\paragraph{Release-related failure modes ($n = 9$)}
These are issues arising due to the \textit{release} of a Generative AI system, which is typically accompanied by documentation or other user-facing communication. These exist largely at the Human interaction layer: that is, they determine whether the Generative AI system ``perform[s] its intended function at the point of use.'' We identify two release-related failure modes: \begin{enumerate*}
    \item \textit{Lack of transparency about model's capabilities/limitations} ($n = 3$), referring to when a system's developer fails to provide clarity about the system's robustness, accuracy, or other performance metric in a manner that is comprehensible to its users, and
    \item \textit{Unanticipated use} ($n = 6$), referring to when a system's developer fails to anticipate a plausible use case of the system.
\end{enumerate*}  

\paragraph{Use-related failure modes ($n = 301$)}
These are issues arising due to the use of the Generative AI system as a result of either user error or user intention. These exist largely at the Human interaction layer. We identify four use-related failure modes: \begin{enumerate*}
    \item \textit{Undisclosed or unwelcome use} ($n = 80$), referring to when the use of Generative AI in a particular context subverts an explicit or implicit expectation of human expertise, labor, or creativity, 
    \item \textit{Contested use} ($n = 7$), referring to when it is unclear whether a Generative AI system was used in a certain context (e.g., resulting in false accusations of 
    dishonesty),
    \item \textit{Improper use} ($n = 23$), referring to when a Generative AI system is used to complete tasks either requiring professional license/training or subject to industry standards and the user fails to properly review outputs, and
    \item \textit{Malicious use} ($n = 198$), referring to when a bad actor uses Generative AI to facilitate harm to others including through the spread of disinformation, fraud, defamation, nonconsensual sexualization, or security threats.
\end{enumerate*} 
\subsubsection{Which failure modes affect each type of stakeholder?}

\begin{figure}[h]
    \centering
    \includegraphics[width=0.7\linewidth]{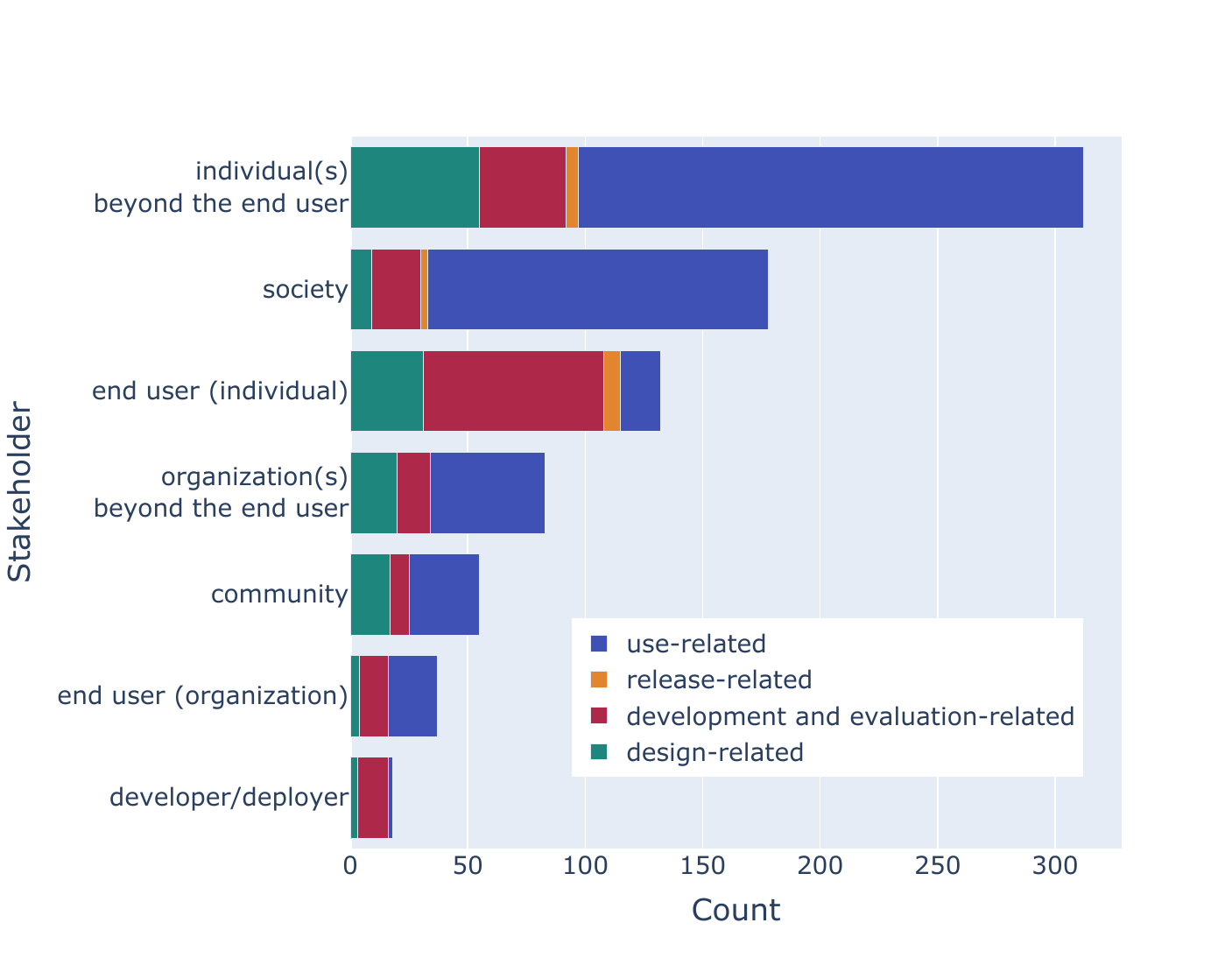}
    \caption{Bars are divided by the type of failure mode that precipitated harm to the respective stakeholder. Use-related failure modes precipitated the vast majority of harms to non-interacting stakeholders while Development and evaluation-related failure modes precipitated most of the harms to interacting stakeholders.}
    \label{fig:entityVcause}
\end{figure}


In Figure \ref{fig:entityVcause}, we report the types of failure modes that contribute to the harms experienced by each type of stakeholder. For all types of non-interacting stakeholders, Use-related failure modes (specifically, Malicious use or Undisclosed or unwelcome use) most often contributed to the harms they experienced. Therefore, not only do most publicly reported incidents involve harms to stakeholders who did not choose to interact with Generative AI and rarely participate in the benefits, but the failure modes most often contributing to those harms are due to user intent. In particular, in cases of Malicious use, the apparent benefit that the Generative AI system offers to its user amounts to intentional harm to other parties. The failure mode that most frequently precipitated harms to interacting stakeholders was Hallucination, a Development and evaluation-related failure mode. 

Thus, safety evaluations at the Capability layer mostly function to mitigate harms to interacting stakeholders while evaluations at the Human interaction layer are necessary (but likely insufficient) to address harms to non-interacting stakeholders.


\subsection{Which failure modes are associated with each type of harm?}


\begin{figure}[h]
    \centering
    \includegraphics[width=0.8\linewidth]{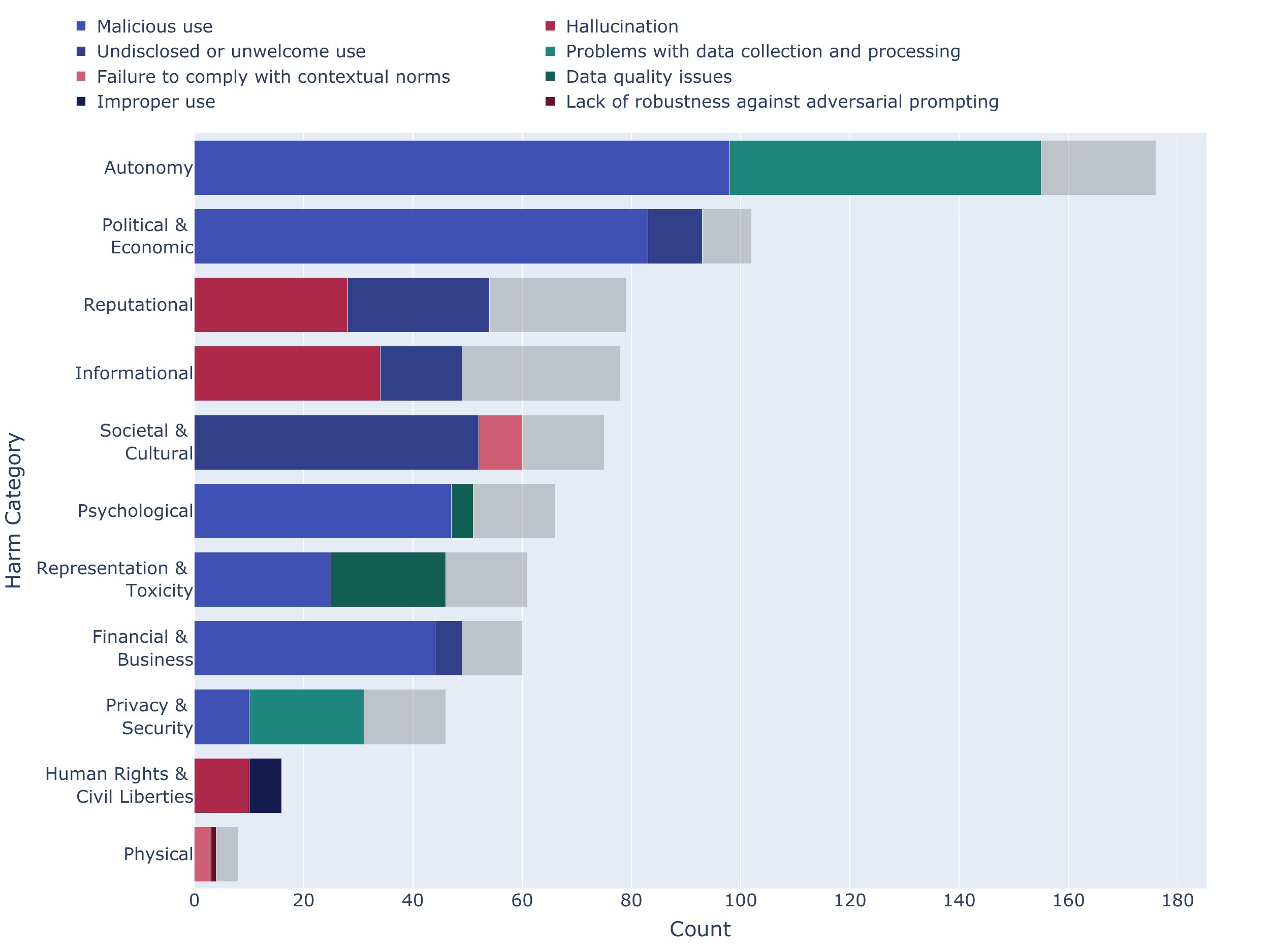}
    \caption{We indicate the two most prevalent sociotechnical failure modes precipitating each category of harm with the colored sections of each bar; the gray sections are all other failure modes combined. We omit Environmental harms due to low frequency in our dataset.}\label{fig:harm_vs_cause}
\end{figure}

As seen in Figure \ref{fig:harm_vs_cause}, some categories of harm are overwhelmingly associated with a single failure mode. 
We examined whether the incidents coded with each of these high-frequency combinations (e.g., Autonomy and Malicious use) converged on particular themes. 

Autonomy, Political \& Economic, Psychological, and Financial \& Business harms were all most often associated with Malicious use. The incidents coded with any of these combinations
were largely precipitated by deepfakes. 
\begin{itemize}
    \item Incidents coded with an Autonomy harm and Malicious use were largely cases of impersonation. For example, several companies were defrauded by deepfake impersonations of their C-suite executives \cite{fraud_AIAAIC0227, fraud_AIAAIC0462} and both public and private individuals were defamed by their impersonations' offensive remarks \cite{jordan_AIAAIC0283, principal_AIAAIC1173}.
    \item Incidents coded with a Political \& Economic harm and Malicious use largely involved deepfakes deployed to create false narratives about political figures \cite{biden_AIAAIC0389, trump_767}, voter groups \cite{trump_650, trump_682}, or current events \cite{putin_AIAAIC1033, pentagon_AIAAIC1019}. 
    \item Incidents coded with a Psychological harm and Malicious use were largely cases where pornographic material of an individual was created without their knowledge \cite{deepfake_496, deepfake_AIAAIC1166}.
    \item Incidents coded with a Financial \& Business harm and Malicious use mostly involved deepfake-facilitated scams \cite{deepfake_AIAAIC1624, deepfake_AIAAIC0213}, fraud \cite{deepfake_AIAAIC1117}, or extortion \cite{deepfake_551, deepfake_AIAAIC0820}.
\end{itemize} 
Incidents coded with Human Rights \& Civil Liberties harms and Hallucination represented, in all but one case, when false information generated by AI was presented in a court of law \cite{legal_AIAAIC1279, legal_709}. Incidents coded with Privacy \& Security harms and Problematic data collection and use mostly involved violations of privacy rights due to Generative AI developers' data acquisition practices, including internet scraping \cite{scrape_AIAAIC1439, scrape_AIAAIC1200}.

Some combinations were more heterogeneous; for example, the combination of Societal \& Cultural harms and Undisclosed or unwelcome use included controversies due to the substitution of AI-generated outputs for human artists \cite{artwork_369, artwork_AIAAIC1136} and students' use of Generative AI to cheat \cite{chatbot_339}. 

\subsubsection{Co-occurring failure modes}
\label{stacking-failures}
As \citet{weidinger_sociotechnical_2023} argue, the Human interaction and Systemic impact layers of the sociotechnical system are critical in determining whether Generative AI harms materialize in the real world. Throughout our analysis, we found that this was demonstrated by several commonly co-occurring failure modes precipitating a particular type of risk. For example, when AI-generated false information was presented in a court of law \cite{legal_AIAAIC1279, legal_709}, we identified both Hallucination and Improper use as associated failure modes -- while the AI system produced the false information (Hallucination), the user was ultimately held at least somewhat responsible for not having verified the AI's output (Improper use). Both failure modes were necessary to have precipitated the harm. 

\subsubsection{Role of open-source models}
\label{open-source}
As noted in prior work, the technical ecosystem that facilitates the creation of AI-generated non-consensual intimate imagery (AIG-NCII) is built upon open-source models \cite{gibson2024analyzingainudificationapplication, ding2025malicioustechnicalecosystemexposing}. Our work both emphasizes the real-world impact of this finding -- our dataset included many incidents of non-consensual deepfake pornography -- and corroborates it: in cases where the nudification tool was known, we found that it was built upon an open-source model such as Stable Diffusion \cite{ly_open_2024, milmo_ai-created_2023,maiberg_samsung-backed_2024}.  

\subsubsection{Effect of output modality} 
\label{modality}
Among the Generative AI systems associated with the incidents in our dataset, \textit{text} was the most prevalent output modality ($n = 219$), followed by \textit{image} ($n = 93$), \textit{video} ($n = 76$), \textit{audio} ($n = 49$), and \textit{multimodal} ($n = 62$).\footnote{Throughout the coding process, we considered \textit{code} a separate modality, but ultimately merged it with text due to a low number of incidents ($n = 3$).} The prevalence of particular sociotechnical failure modes and harm categories was highly dependent on the output modality of the Generative AI system at fault.
Hallucination was a prevalent failure mode for systems producing text while Malicious use was the most prevalent failure mode for systems producing any other output modality. 

\section{Discussion} \label{discussion}

Our findings offer a quantitative overview of the harms and sociotechnical failure modes of Generative AI systems deployed in the real world. 
Here, we discuss the benefits of eliciting not only the harms, but also \textit{who} they affect and the \textit{sociotechnical failures} that underlie them when conducting case studies of real-world incidents. We conclude with several recommendations for key AI stakeholders.


\subsection{Implications of Our Findings}
Previous taxonomies of AI and Generative AI harms have identified, categorized, and suggested the implications of a wide range of harms \cite{abercrombie_collaborative_2024, weidinger_sociotechnical_2023, hutiri_not_2024, zeng_ai_2024, shelby_sociotechnical_2023, critch_tasra_2023, solaiman_evaluating_2024, liu_trustworthy_2024, bird_typology_2023, barnett_ethical_2023, vidgen_introducing_2024, hoffman, pittaras_taxonomic_nodate, slattery_ai_nodate}. Separately, work has been done to identify \textit{failure modes} of AI~\cite{raji_fallacy_2022,slattery_ai_nodate}.  
However, most previous work lacks a real-world understanding of how these harms and failure modes relate to each other. In this paper, we develop a \textit{relational} understanding of harms, failure modes, and affected stakeholders. 
This coordinated approach lends itself well to mapping pathways of accountability and identifying where they break down, helping inform which stakeholders are best positioned to address each harm or risk.


\subsubsection{The uneven distribution of benefits and harms of Generative AI across stakeholders erodes pathways of accountability.}
\label{uneven-distribution}
In particular, our findings suggest that the balance of benefits and harms of Generative AI differs greatly between its interacting and non-interacting stakeholders. We find that the touted benefits of Generative AI (e.g., increased productivity) accrue mostly to its interacting stakeholders while its non-interacting stakeholders bear many of the most critical and prevalent harms, such as harms to their autonomy, psychological wellbeing, and civil liberties. This has two main implications. First, interacting stakeholders tend not to need to assume substantive risk of harm for their choice to participate in the benefits of Generative AI, eroding one path of accountability. Second, ethical and safety considerations at the Human interaction and Systemic impact layers of Generative AI sociotechnical systems are highly necessary; however, because non-interacting stakeholders affected by failures at these layers often must rely on indirect and inefficient pathways of recourse -- for instance, those harmed by the copyright issues resulting from Generative AI developers' data acquisition practices are suing the relevant developers \cite{times_AIAAIC1264, voice_AIAAIC1431, sue_AIAAIC1234} -- developers are rarely incentivized to conduct them. Thus, ethics and safety improvements at these layers likely must be either incentivized or enforced by external oversight bodies.  
We see this in other domains where this lopsided distribution of benefits and risks is present, such as data privacy, where those who collect data benefit while assuming little risk, and those whose data are collected rarely benefit but assume substantive risk.

Certain cumulative harms, which, as noted in Section \ref{limitations}, are unlikely to surface in analyses of publicly reported incidents may accrue largely to the end users of Generative AI. This may change the balance of harms and benefits that end users assume over time. These harms include the deterioration of critical thinking skills due to over-reliance on Generative AI systems, psychological addiction, and psychological isolation \cite{abercrombie_collaborative_2024}.

\subsubsection{Generative AI changes the landscape of AI harms.} 
As we discuss in Section \ref{how-what-who}, \citet{velazquez_decoding_2024} applied a framework similar to ours in their analysis of 639 AI incident reports. While their analysis includes incidents involving a wide range of AI systems, we focused our analysis strictly on incidents involving \textit{Generative} AI. Their work shares some key findings with this paper: for example, they found that Autonomy harms were the most prevalent type of harm in their dataset, which we also found (Figure \ref{fig:risk_tax}). However, they also came to several conclusions that contrast with our findings. For example, they found that the vast majority of incidents in their dataset harmed the individual directly interacting with the AI system, while we found the opposite: the overwhelming majority of incidents in our dataset harmed stakeholders that did \textit{not} directly interact with the Generative AI system (Figure \ref{fig:risk_cats}). They also found that most harms were non-malicious while Malicious use was the single most prevalent failure mode in our analysis (Figure \ref{fig:failure_tax}). We argue that these contrasting findings are due to differences between how \textit{Generative} AI and traditional AI harms manifest in the real world; as noted in Section \ref{how-what-who}, this argument is corroborated by the findings of \citet{hutiri_not_2024}.

These differences, in concert with our finding that Generative AI systems with different output modalities tend to produce different types of harm via different failure modes (Section \ref{modality}), suggest that effective AI governance is necessarily application-specific.

\subsection{Recommendations}
As argued in Section \ref{uneven-distribution}, the distribution of benefits and harms of Generative AI across stakeholders creates an incentive structure that leads to the neglect of non-interacting stakeholders. Therefore, we call on academic, industry, legal, and policy actors to prioritize and advocate for non-interacting stakeholders' interests. We outline three recommendations that aim to address some of the most prevalent and/or severe harms in our analysis, including Autonomy, Psychological, and Human Rights \& Civil Liberties harms.

\subsubsection{Basic AI literacy can reduce certain risks of harm} Harm is often precipitated by not just one, but two or more failures in combination (Section \ref{stacking-failures}). This suggests that preventing just one of the precipitating failures may have prevented these harms altogether. Capitalizing on this insight, we advocate for work toward promoting basic AI literacy in the general public, which may help minimize Use-related failures. In particular, several incidents precipitated by Improper use demonstrate that members of the general public do not understand the capabilities, limitations, or risks of Generative AI. For example, when a Canadian lawyer cited hallucinations in a divorce case, she claimed that she did not realize that ChatGPT could hallucinate \cite{canada_AIAAIC1371}.

Because Generative AI applications are often marketed 
as consumer products, the general public is a critical audience and participant in conversations about Generative AI governance. Yet, many members of the public lack the knowledge and vocabulary to describe the behavior of Generative AI applications.
We argue that it is critical to communicate current understandings of the capabilities, limitations, and harms of Generative AI in ways that are comprehensible to the general public. Some work is being done on consumer-facing \textit{safety labels} \cite{chia_safety_2024}, which aim to encourage developers and deployers to ``be transparent with users by providing information on how the generative AI models and apps work.'' 
This is an important step, but 
it is insufficient: many of the harms of Generative AI affect those who do \textit{not} interact directly with it, and hence are unlikely to see these labels. Taxonomies of harms and failure modes are important artifacts for the general public, but are currently largely directed toward expert audiences. Hence, an important area of future research is to create publicly accessible and digestible information about Generative AI failure modes and harms. In the near future, we aim to distill both taxonomies developed in this paper into a non-technical and concise document and assess its usability for members of the public.



\subsubsection{The current regulatory landscape enables the development of Generative AI tools specifically for abusive purposes, especially for the creation of AIG-NCII} The EU AI Act incentivizes open-source development by reducing the associated obligations of documentation and disclosure \cite{eu_ai}. As \citet{kapoor} caution, however, while open-source models offer benefits such as increasing innovation, they also pose greater risks of misuse. Our findings corroborate this concern (Section \ref{open-source}). Although we acknowledge that, when limited to publicly available data, it is difficult to confidently determine whether an incident involved an open or closed model, we conjecture that the majority of egregious applications (for example, websites enabling the creation of AIG-NCII) are built on open-source models. This suggests that more research centered on real-world evidence of the supposed risks and benefits of open-source foundation models is necessary to inform regulatory stances. 

We argue that the current regulatory landscape enables the creation of AIG-NCII. While there has been some work to regulate the spread of AIG-NCII, such as the Take It Down Act \cite{takeitdown}, these measures still place the burden of removal on the victims \cite{ding2025malicioustechnicalecosystemexposing}. Therefore, minimizing the \textit{creation} of AIG-NCII is critical to minimize its psychological harms. \citet{gibson2024analyzingainudificationapplication} surface several viable solutions, including, as we have discussed, critical analysis of the benefits of open-source models, as well as consideration of access restrictions and tracking of downstream usage. They also note that in the absence of any regulation requiring verification of deepfake subjects' consent or age, many nudification tools do not even mention these issues. Overall, the current lack of regulation of the open-source technical ecosystem enables the existence of tools specifically positioned to be abusive. 





\subsubsection{Provenance data tracking and synthetic content detection can reduce some risks of harm} The NIST AI 100-4 \cite{national_institute_of_standards_and_technology_us_reducing_2024} proposes provenance data tracking and synthetic content detection as strategies to reduce risks posed by synthetic content, including AIG-NCII and misinformation.
However, some researchers find that these solutions are incomplete or misaligned. \citet{ding2025malicioustechnicalecosystemexposing} argue that just because AIG-NCII is detectable does not mean it is harmless, and \citet{kapoor_we_2025} argue that the real bottleneck for misinformation is \textit{distribution} rather than creation -- thus, fears surrounding the proliferation of AI-generated misinformation were overblown. 

Nonetheless, we argue that data tracking and synthetic content detection can reduce some risks from synthetic content. In particular, researchers have cautioned that Generative AI could still meaningfully change the landscape of online misinformation, although in more incremental ways than originally anticipated \cite{brookings_misunderstood}. We corroborate this argument: we found several incidents where Generative AI was deployed to generate convincing misinformation, causing significant harm. For example, an AI-generated image of an explosion at the Pentagon sparked a dip in the US stock market \cite{bond_fake_2023}. In addition, we surfaced many incidents of deepfake-facilitated scams \cite{deepfake_AIAAIC1624, deepfake_AIAAIC0213}, fraud \cite{deepfake_AIAAIC1117}, and extortion \cite{deepfake_551, deepfake_AIAAIC0820}. These incidents have caused significant financial and psychological harm to their victims, and we argue that they can be effectively mitigated by data tracking and synthetic content detection. 

\section{Conclusion} \label{conclusion}
In this paper, we set out to develop an understanding of real-world Generative AI harms, the sociotechnical failures that precipitate them, and who they affect. We ground our work in a systematic review of 499 publicly reported Generative AI incidents. We find that many incidents result in harm to those who do \textit{not} interact directly with Generative AI (such as the unknowing subject of a deepfake), but are often precipitated by those who \textit{do} interact directly with Generative AI (the end users and developers). Our findings also suggest that the landscape of Generative AI harms is meaningfully different from that of traditional AI systems: for example, malicious use may be a far more prevalent failure mode for Generative AI systems than for traditional AI systems. Given these insights, we make several recommendations for key AI stakeholder groups to prioritize and mitigate Generative AI harms, including through public disclosures and education, regulatory stances on open-source development, and provenance data tracking and synthetic content detection.


\begin{acks}
    This work was supported in part by Meta, a Carnegie Mellon University Rales Fellowship, and the Block Center for Technology and Society at Carnegie Mellon University.
\end{acks}

\bibliographystyle{ACM-Reference-Format}
\bibliography{refs}

\clearpage

\section{Appendices}

\subsection{Overview of Existing AI Risk and Harm Taxonomies}
We include a nonexhaustive overview of existing AI risk and harm taxonomies on the next page. 

\subsection{Codebook of Harms}  \label{appendix_harm}
Our codebook of harms is grounded in prior literature, including \citet{abercrombie_collaborative_2024} (the \textit{Autonomy, Political \& Economic, Societal \& Cultural, Reputational, Psychological, Financial \& Business, Human Rights \& Civil Liberties, Physical,} and \textit{Environmental} categories),  \citet{weidinger_sociotechnical_2023} (the \textit{Misinformation} and \textit{Representation \& Toxicity} categories), and \citet{lee_deepfakes_2024} (the \textit{Privacy \& Security} category).

Our complete codebook of harms is available in the ``Harm Codebook" tab \href{https://docs.google.com/spreadsheets/d/1eB0hdhSwQQQNMO9gnqGPWKZnhhCypvz6swqcEbxJdgU/edit?usp=sharing}{here}.


\subsection{Codebook of Sociotechnical Failure Modes}  \label{appendix_failures}
Our complete codebook of sociotechnical failure modes is available in the ``Sociotechnical Failure Mode Codebook" tab \href{https://docs.google.com/spreadsheets/d/1eB0hdhSwQQQNMO9gnqGPWKZnhhCypvz6swqcEbxJdgU/edit?usp=sharing}{here}.

\subsection{Codebook of Affected Stakeholders}  \label{appendix_entities}
Our complete codebook of affected stakeholders is available in the ``Stakeholder Codebook" tab \href{https://docs.google.com/spreadsheets/d/1eB0hdhSwQQQNMO9gnqGPWKZnhhCypvz6swqcEbxJdgU/edit?usp=sharing}{here}.

\subsection{Coded Incident Dataset}  \label{appendix_incidents}
Our complete coded incident dataset is available in the ``Coded Dataset" tab \href{https://docs.google.com/spreadsheets/d/1eB0hdhSwQQQNMO9gnqGPWKZnhhCypvz6swqcEbxJdgU/edit?usp=sharing}{here}. 

\begin{table*}[h!]
    \centering
    \caption{Overview of Existing AI Risk and Harm Taxonomies}
    \begin{tabular}{p{1.1in} p{1in} p{1.5in} p{2.5in}}\hline
        \textbf{Taxonomy} & \textbf{Technology} & \textbf{Target Audience} & \textbf{Methodology} \\\midrule
        AIAAIC \cite{abercrombie_collaborative_2024} & AI and automation & Broad, including the general public & Use case analysis, literature review, expert outreach, annotation testing on incident reports ($n$ = 39), validation through broader community review \\ 
        AIID \cite{pittaras_taxonomic_nodate} & AI & Technical (AI experts) & Incident annotation using proposed three-part framework ($n$ = 41)  \\ 
        CSET \cite{hoffman} & AI & AI community & Discussions with outside organizations, annotating incidents in the AIID ($n\approx$ 100)	\\  \citet{shelby_sociotechnical_2023} & Algorithmic Systems & Practitioners and researchers & Literature review ($n$ = 172) \\ 
        \citet{lee_deepfakes_2024} & AI & Researchers, practitioners & Systematic review of privacy-related AIAAIC incidents ($n$ = 321) \\ 
        TASRA \cite{critch_tasra_2023} & AI & AI community & Literature review, decision tree analysis \\ 
        AIR \cite{zeng_ai_2024} & AI & Researchers, evaluators & Analysis of 8 government policies and 16 company policies \\ \hline
        \citet{vidgen_introducing_2024} & General-purpose AI chatbots & AI safety researchers and practitioners & Expert interviews, literature review, analysis of safety evaluation datasets, policy papers, and community guidelines from industry \\ 
        \citet{weidinger_sociotechnical_2023} & Generative AI & Researchers, practitioners, evaluators & Literature review \\
        \citet{barnett_ethical_2023}& Generative audio models & Researchers & Literature review \\ 
        \citet{bird_typology_2023} & Generative text-to-image models & Researchers, practitioners, regulators & Literature review \\ 
        \citet{liu_trustworthy_2024}& LLMs & Researchers, developers, evaluators & Literature review, researchers’ expertise \\ 
        \citet{solaiman_evaluating_2024} & Generative AI & Researchers, developers, auditors, policymakers & Expert workshops \\ 
        \citet{hutiri_not_2024}& Speech generation models & Policymakers, developers, researchers & Iterative design approach through incident annotation ($n$ = 35) from AIID, AIAAIC, and OECD \\
    \end{tabular}
    \label{tab:tax_overview}
\end{table*}

\end{document}